\documentclass[pre,twocolumn,showpacs,aps,floats,floatfix,superscriptaddress]{revtex4}

\usepackage{graphicx}
\usepackage{amsmath}

\begin{document}

\newcommand{\avk}{\langle k \rangle}
\newcommand{\fluck}{\langle k^2 \rangle}
\newcommand{\G}{\mathcal{G}}
\newcommand{\V}{\mathcal{V}}
\newcommand{\E}{\mathcal{E}}
\newcommand{\C}{\mathcal{C}}
\newcommand{\bx}{\mathbf{x}}
\newcommand{\by}{\mathbf{y}}
\newcommand{\ie}{\emph{i.e.}}
\newcommand{\abs}[1]{\left| #1 \right|}
\newcommand\Erdos{Erd\H{o}s}
\newcommand\ER{Erd\H{o}s-Renyi{}}
\newcommand\BA{Barab{\'a}si-Albert{}}

\title{Co-Betweenness: A Pairwise Notion of Centrality}

\author{Eric D. Kolaczyk}
\affiliation{Dept. of Mathematics and Statistics\\Boston University\\
Boston, MA,  USA}
\author{David B. Chua}
\affiliation{Boston, MA}
\author{Marc Barth\'elemy}
\affiliation{CEA-Centre d'Etudes de
Bruy{\`e}res-le-Ch{\^a}tel, D\'epartement de Physique Th\'eorique et
Appliqu\'ee BP12, 91680 Bruy\`eres-Le-Ch\^atel, France}

\date{\today} 
\widetext

\begin{abstract}
  
  Betweenness centrality is a metric that seeks to quantify a sense of
  the importance of a vertex in a network graph in terms of its
  `control' on the distribution of information along geodesic paths
  throughout that network. This quantity however does not capture
  how different vertices participate {\em together} in such control.
  In order to allow for the uncovering of finer details in this regard,
  we introduce here an extension of
  betweenness centrality to pairs of vertices, which we term {\em
    co-betweenness}, that provides the basis for quantifying various
  analogous pairwise notions of importance and control.  More
  specifically, we motivate and define a precise notion of
  co-betweenness, we present an efficient algorithm for its
  computation, extending the algorithm of~\cite{brandes01} in a
  natural manner, and we illustrate the utilization of this
  co-betweenness on a handful of different communication networks. From
  these real-world examples, we show that the co-betweenness allows one
  to identify certain vertices which are not the most central vertices but
  which, nevertheless, act as important actors in the relaying and
  dispatching of information in the network.

\end{abstract}



\maketitle 


\section{Introduction}
\label{sec:introduction}

In social network analysis, the problem of determining the importance
of actors in a network has been studied for a long time (see, for
example, \cite{Wasserman:1994}). It is in this context that the
concept of the {\em centrality} of a vertex in a network emerged.
There are numerous measures that have been proposed to numerically
quantify centrality which differ both in the nature of the underlying
notion of vertex importance that they seek to capture, and in the
manner in which that notion is encoded through some functional of the
network graph. See~\cite{borgatti.everett}, for example, for a
recent review and categorization of centrality measures.

Paths -- as the routes by which flows (e.g., of information or
commodities) travel over a network -- are fundamental to the
functioning of many networks.  Therefore, not surprisingly, a number
of centrality measures quantity importance with respect to the
sharing of paths in the network.  One popular measure is
\emph{betweenness centrality}.  First introduced in its modern form
by~\cite{freeman77}, the betweenness centrality is essentially a
measure of how many geodesic (ie., shortest) paths run over a given
vertex.  In other words, in a social network for example, the
betweenness centrality measures the extent to which an actor ``lies
between'' other individuals in the network, with respect to the
network path structure.  As such, it is a measure of the control that
actor has over the distribution of information in the network.

The betweenness centrality -- as with all other centrality measures of
which we are aware -- is defined specifically with respect to a single
given vertex.  In particular, vertex centralities produce
an ordering of the vertices in terms of their individual importance,
but do not provide insight into the manner in which vertices act
together in the spread of information across the network.  Insight
of this kind can be important in presenting an appropriately more
nuanced view of the roles of the different vertices, beyond their
individual importance.  A first natural extension of the
idea of centrality in this manner is to pairs of vertices.

In this paper, we introduce such an extension, which we term the
\emph{co-betweenness centrality}, or simply the {\em co-betweenness}.
The co-betweenness of two vertices is essentially a measure of how
many geodesic paths are shared by the vertices, and as such provides
us with a sense of the interplay of vertices across the network.  For
example, the co-betweenness alone quantifies the extent to which pairs
of vertices jointly control the distribution of information in the
network.  Alternatively, a standardized version of co-betweenness
produces a well-defined measure of correlation between flows over the
two vertices.  Finally, an alternative normalization quantifies the
extent to which one vertex controls the distribution of information to
another vertex.

This paper is organized as follows.  In Section~\ref{sec:bg}, we briefly
review necessary technical background.  In Section~\ref{sec:cob},
we provide a precise definition for the co-betweenness and related measures,
and motivate each in the context of an Internet communication network.  
An algorithm for the efficient computation of co-betweenness, for all pairs
of vertices in a network, is sketched in Section~\ref{sec:computation}, 
and its properties are discussed.  In Section~\ref{sec:applications}, we 
further illustrate our measures using two social networks whose ties are 
reflective of communication.  Some additional discussion is provided in
Section~\ref{sec:disc}.  Finally, a formal description of our algorithm,
as well as pseudo-code, may be found in the appendix.

\section{Background}
\label{sec:bg}

Let $\G=(\V,\E)$ denote an undirected, connected network graph with
$n_v$ vertices in $\V$ and $n_e$ edges in $\E$.  A \emph{walk} on
$\G$, from a vertex $v_0$ to another vertex $v_{\ell}$, is an
alternating sequence of vertices and edges, say
$\{v_0,e_1,v_1,\ldots,v_{\ell-1},e_{\ell},v_{\ell}\}$, where the
endpoints of $e_i$ are $\{v_{i-1},v_i\}$.  The {\it length} of this
walk is said to be $\ell$.  A \emph{trail} is a walk without repeated
edges, and a \emph{path}, a trail without repeated vertices.  A
shortest path between two vertices $u, v\in \V$ is a path between $u$
and $v$ whose length $\ell$ is a minimum.
Such a path is also called a {\em geodesic}
and its length, the {\em geodesic distance} between $u$ and $v$.  In
the case that the graph ${\cal G}$ is weighted i.e., there is a collection
of edge weights $\{w_e\}_{e\in\E}$, where $w_e\ge 0$, shortest paths
may be instead defined as paths for which the total sum of edge
weights is a minimum.  In the material that follows, we will restrict
our exposition primarily to the case of unweighted graphs, but
extensions to weighted graphs are straightforward.  For additional
background of this type, see, for example, the 
textbook~\cite{Clark:1991}.

Let $\sigma_{s t}$ denote the total number of shortest paths that
connect vertices $s$ and $t$ (with $\sigma_{s s} \equiv 1$), and let
$\sigma_{s t}(v)$ denote the number of shortest paths between $s$ and
$t$ that also run over vertex $v$.  Then we define the betweenness
centrality of a vertex $v$ as a weighted sum of the number of paths
through $v$,
\begin{equation}
  \label{eq:45}
  B(v) = \sum_{s,t\in \V \setminus\{ v \}}
  \frac{\sigma_{s t}(v)}{\sigma_{s t}} \text{.}
\end{equation}
Note that this definition excludes the shortest paths that start
or end at $v$.  However, in a
connected graph we will have $\sigma_{s t}(v) = \sigma_{s t}$
whenever $s=v$ or $t=v$, so the exclusion amounts to removing a
constant term that would otherwise be present in the betweenness
centrality of every vertex.

\begin{figure}  
  \centerline{
  \includegraphics[angle=0,scale=0.4]{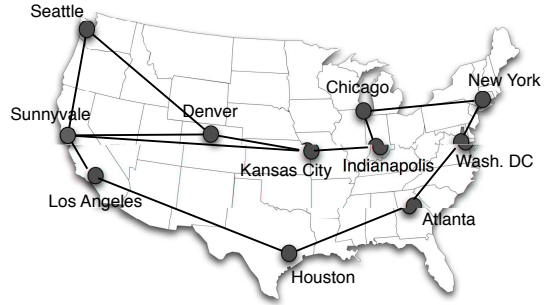}
	}
  \caption{Graph representation 
	of the physical topology of the Abilene network.  Nodes represent
	regional network aggregation points (so-called `Points-of-Presence'
	or PoP's), and are labeled according to their metropolitan
	area, while the edges represent systems of optical 
	transportation technologies and routing devices.}
  \label{fig:abilene.network}
\end{figure}

As an illustration, which we will use throughout this section and the 
next, consider the network in Figure~\ref{fig:abilene.network}.
This is the Abilene network, an Internet network that is part of the
Internet2 project \cite{address0}, a research
project devoted to development of the `next generation' Internet.  It
serves as a so-called `backbone' network for universities and research
labs across the United States, in a manner analogous to the federal 
highway system of roads.  We use this network for illustration
because, as a technological communication network, the notions of
connectivity, information, flows, and paths are all explicit and physical,
and hence facilitate our initial discussion of betweenness and
co-betweenness.  Later, in Section~\ref{sec:applications}, we will
illustrate further with two communication networks from the social
network literature.

The information traversing this network takes the form of so-called
`packets', and the packets flow between origins and destinations on
this network along paths strictly determined according to a set of
underlying routing protocols (Technically, the Abilene network
  is more accurately described by a directed graph.  But, given the
  fact that routing is typically symmetric in this network, we follow
  the Internet2 convention of displaying Abilene using an undirected
  graph.).  A reasonable first approximation of the routing of
information in this network is with respect to a set of unique
shortest paths.  In this case, the betweenness $B(v)$ of any given
vertex $v\in\V$ will be exactly equal to the number of shortest paths
through $v$.  The vertices in Figure~\ref{fig:abilene.network}
correspond to metropolitan regions, and have been laid out roughly
with respect to their true geographical locations.  Intuitively and
according to earlier work on centrality in spatial
networks~\cite{Barrat:2005}, one might suspect that vertices near the
central portion of the network, such as Denver or Indianapolis, have
larger betweenness, being likely forced to support most of the flows
of communication between east and west.  We will see in
Section~\ref{sec:cob} that such is indeed the case.

Until recently, standard algorithms for computing betweenness
centralities $B(v)$ for all vertices in a network had $O(n_v^3)$
running times, which was a stumbling block to their application in
large-scale network analyses. Faster algorithms now exist, such as
those introduced in \cite{brandes01}, which have running time of
$O(n_v n_e)$ on unweighted networks and $O(n_v n_e+n_v^2\log n_v)$ on
weighted networks, with an $O(n_v+n_e)$ space requirement.  These
improvements derive from exploiting a clever recursive relation for
the partial sums $\sum_{t\in\V} \sigma_{s,t}(v)/\sigma_{s,t}$.  As we
will see, the need for efficient algorithms is even more important in
the case of the co-betweenness, and we will make similar usage of
recursions in developing an efficient algorithm for computing this
quantity.

\section{Co-Betweenness}
\label{sec:cob}

We extend the concept of vertex betweenness centrality to pairs of
vertices $u$ and $v$ by letting $\sigma_{s t}(u,v)$ denote the
number of shortest paths between vertices $s$ and $t$ that pass
through both $u$ and $v$, and defining the vertex co-betweenness
as
\begin{equation}
  \label{eq:CoB}
  \C(u,v) = \sum_{\substack{s,t \in \V \setminus \{u,v\}}}
  \frac{\sigma_{s t}(u,v)}{\sigma_{s t}} \text{\;.}
\end{equation}
Thus co-betweenness gives us a measure of the number of shortest
paths that run through both vertices $u$ and $v$. 

To gain some insight into the relation between betweenness and
co-betweenness, consider the following statistical perspective.
Recall the Abilene network described in the previous section, and
suppose that $x_{s,t}$ is a measure of the information (i.e., Internet
packets) flowing between vertices $s$ and $t$ in the network.
Similarly, let $y_v$ be the total information flowing through vertex
$v$.  Next, define $\bx$ to be the $n_p\times 1$ vector of values
$x_{s,t}$, where $n_p$ is the total number of pairs of vertices
exchanging information, and $\by$, to be the $n_v\times 1$ vector of
values $y_v$.  A common expression modeling the relation between these
two quantities is simply $\by = R\bx$, where $R$ is an $n_v\times n_p$
matrix (i.e., the so-called `routing matrix') of $0$'s and $1$'s,
indicating through which vertices each given routed path goes.

Now if $\bx$ is considered as a random variable, with uncorrelated elements,
then its covariance matrix is simply equal to the $n_p\times n_p$
identity matrix.  The elements of $\by$, however, will be correlated,
and their covariance matrix takes the form $\Omega=RR^{T}$, by virtue of
the linear relation between $\by$ and $\bx$.  Importantly, note that 
the diagonal elements of $\Omega$ are the betweenness' $B(v)$.  
Furthermore, the off-diagonal elements are the co-betweenness'
$\C(u,v)$.  When shortest paths are not unique, the same results hold
if the matrix $R$ is expanded so that each shortest path between a
pair of vertices $s$ and $t$ is afforded a separate column, and the
non-zero entries of each such column has the value $\sigma_{s,t}^{-1}$,
rather than $1$.  In this case, $R$ may be interpreted as a stochastic
routing matrix.

\begin{figure}   
  \centerline{
  \includegraphics[angle=0,scale=0.40]{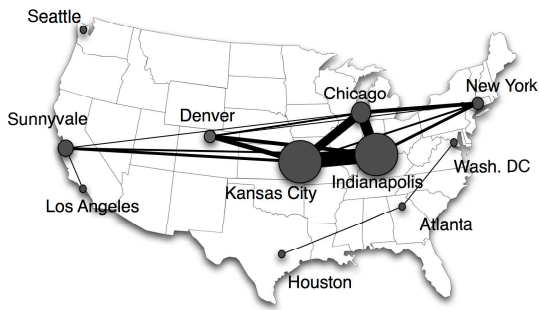}
  }
  \caption{Graph representation of the betweenness and
	co-betweenness values for the Abilene network.  Vertices are
	in proportion to their betweeness.  The width of each link
	is drawn in proportion to the co-betweenness of the two
	vertices incident to it.}
  \label{fig:abilene.CoB}
\end{figure}

To illustrate, in Figure~\ref{fig:abilene.CoB}, 
we show a network graph representation of 
the matrix $\Omega$ for the Abilene network.
The vertices are again placed roughly with respect to their
actual geographic location, but are now drawn in proportion
to their betweenness.  Edges between pairs of vertices now represent
non-zero co-betweenness for the pair, and are also drawn with a
thickness in proportion to their value.  A number of interesting
features are evident from this graph.  First, we see that, as 
surmised earlier, the more centrally located vertices tend to have the
largest betweenness values.  And it is these vertices that typically
are involved with the larger co-betweenness values.  Since the
paths going through both a vertex $s$ and a vertex $t$ are a subset
of the paths going through either one or the other, this tendancy
for large co-betweenness to associate with large betweenness
should not be a surprise.  Also note that the co-betweenness
values tend to be smaller between vertices separated by a larger
geographical distance, which again seems intuitive.

Somewhat more surprising perhaps, however,
is the manner in which the network becomes disconnected.  The 
Seattle vertex is now isolated, as there are no paths
that route through that vertex -- only to and from.  Additionally,
the vertices Houston, Atlanta, and
Washington now form a separate component in this graph,
indicating that information is routed on paths running 
through both the first two and the last two, but not through all three,
and also not through any of these and some other vertex.  Overall,
one gets the impression of information being routed primarily 
over paths along the upper portion of the network in 
Figure~\ref{fig:abilene.network}.  A similar observation has been 
made in~\cite{chua05:kriging}, using different techniques.

While the raw co-betweenness values appear to be quite informative,
one can imagine contexts in which it would be useful to compare
co-betweenness' across pairs of vertices in a manner that adjusts
for the unequal betweenness of the participating vertices.  The
value
\begin{equation}
\C^{corr}(u,v) = \frac{\C(u,v)}{\sqrt{B(u)B(v)}}
\label{eq:CoB.corr}
\end{equation}
is a natural candidate for a standardized version of the
co-betweeness in (\ref{eq:CoB}), being simply the corresponding
entry of the correlation matrix deriving from $\Omega=RR^T$.

\begin{figure}   
  \centering
  \includegraphics[angle=0,scale=0.40]{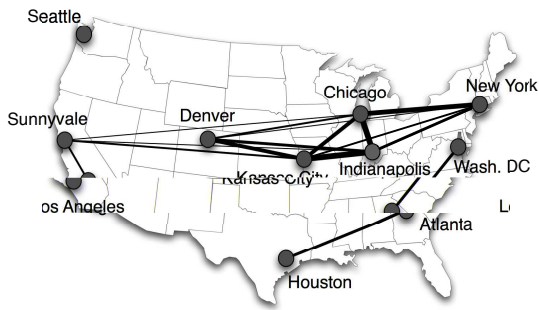}
        \caption{Graph representation of the standardized 
        co-betweenness values $\C^{corr}$ for the Abilene network.
	Vertices are all drawn with equal size.  Edge width is
	drawn in proportion to the standardized co-betweenness
	of the two vertices indicent to it.}
  \label{fig:abilene.CoB.corr}
\end{figure}

Figure~\ref{fig:abilene.CoB.corr} shows a network graph 
representation of the quantities in $\C^{corr}$ for the Abilene
network, with edges again drawn in proportion to the values and 
vertices now naturally all drawn to be the same size.
Much of this network looks like that in 
Figure~\ref{fig:abilene.CoB}.  The one notable exception is that
the magnitude of the values between the three vertices in
the lower subgraph component are now of a similar order to 
most of the other values in the other component.  This fact may be
interpreted as indicating that among themselves, adjusting for
the lower levels of information flowing through this part of the
network, these vertices are as strongly `correlated' as many of
the others.

The co-betweenness may also be used to define a directed notion
of the strength of pairwise relationships.  Let
\begin{equation}
\C(u|v) = \frac{\C(u,v)}{B(v)}
\label{eq:CoB.condp}
\end{equation}
denote the relative proportion of shortest paths through $v$ that also 
go through $u$.  This quantity may be interpreted as a measure
of the control that vertex $v$ has over the information
that passes through vertex $u$.  Alternatively, under uniqueness of
shortest paths, if from among the set of shortest paths through $v$
one is chosen uniformly at random, 
the value $\C(u|v)$ is the probabilty that the chosen path will
also go through $u$.  We call $\C(u|v)$ the {\em conditional
betweenness} of $u$, given $v$.
Note that, in general, $\C(u|v)\ne \C(v|u)$.

\begin{figure}   
  \centering
  \includegraphics[angle=0,scale=0.40]{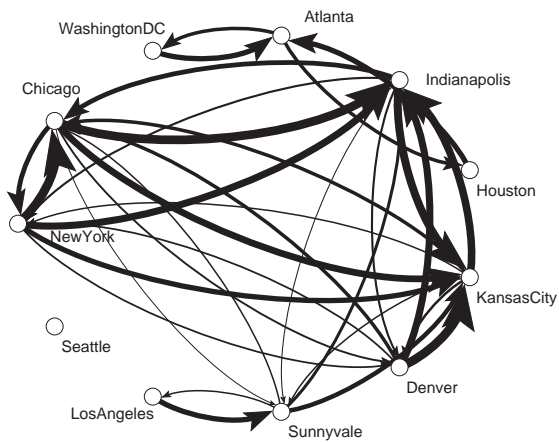}
  \caption{Directed graph representation of the conditional
    betweenness values $\C(u|v)$ (given by Eq.~(4)) for the Abilene
    network. Edges are drawn with width in proportion 
    to their value of $\C(u|v)$ 
    and indicate how one vertex (at the head) controls the 
    flow of information through another (at the tail).}
  \label{fig:abilene.CoB.condp}
\end{figure}

Figure~\ref{fig:abilene.CoB.condp} shows a graph representation
of the values $\C(u|v)$ for the Abilene network.
Due to the asymmetry of these values in $u$ and $v$, arcs are used,
rather than edges, with an arc from $v$ to $u$ corresponding to
$C(u|v)$.  The thickness of the arcs is proportional to these values,
and is therefore indicative of the control exercised on the vertex at
the tail by the vertex at the head.  For improved visualization, we
have used a simple circular layout for the vertices.  Examination of
this figure shows symmetry in the relationships between some pairs of
vertices, but a strong asymmetry between most others.  For example,
vertices like Indianapolis, which were seen previously to have a large
betweenness, clearly exercise a strong degree of control over almost
any other vertices with which they share paths.  More interestingly,
note that certain vertices that are neighbors in the original Abilene
network have more symmetric relationships than others.  The
conditional betweenness' for Atlanta and Washington, DC, are fairly
similar in magnitude, while those for Los Angeles and Sunnyvale are
quite dissimilar, with the latter evidently exercising a noticeably
greater degree of control over the former.

\section{Computation of Co-Betweenness}
\label{sec:computation}

We discuss here the calculation of the co-betweenness values
$\C(u,v)$ in (\ref{eq:CoB}), for all pairs $(u,v)$, from which
the other quantities in (\ref{eq:CoB.corr}) and (\ref{eq:CoB.condp})
follow trivially.
At a first glance, it would appear that an algorithm of 
$O(n_v^4)$ running time is necessary, given that the number of vertex
pairs grows as the square of the number of vertices.  Such an
implementation would render the notion of co-betweenness infeasible
to implement in any but network graphs of relatively modest size.
However, exploiting ideas similar to those underlying the algorithms 
of~\cite{brandes01} for calculating the betweenness' $B(v)$, 
a decidedly more efficient implementation may be obtained,
as we now describe briefly.  Details may be found in the appendix.

Our algorithm for computing co-betweenness involves a three-stage
procedure for each vertex $v\in\V$.  In the first stage, we perform a
breadth-first traversal of the network graph $\G$, to quickly compute
intermediary quantities such as $\sigma_{s v}$, the number of shortest
paths from a source $s$ to each other vertex $v$ in the network; in
the process we form a directed acyclic graph that contains all
shortest paths leading from vertex $s$.
In the second stage, we iterate through each vertex in order of
decreasing distance from $s$ and compute a score $\delta_s(v)$ for
each vertex that is related to its contribution to the
co-betweenness. These contributions are then aggregated in a
depth-first traversal of the directed acyclic graph, which is
carried out in the third and final stage.

In order to compute the number of shortest paths $\sigma_{s v}$ in
the first stage, we note that the number of shortest paths from
$s$ to a vertex $v$ is the sum of all shortest paths to each parent
of $v$ in the directed acyclic graph rooted at $s$, namely,
\begin{equation}
  \label{eq:61}
  \sigma_{s v} = \sum_{t\in p_s(v)} \sigma_{s t} \text{.}
\end{equation}
In the case of an undirected graph, this can be computed in the
course of a breadth-first search with a running time of $O(n_e)$.

In the second stage, we compute $\delta_s(v)$
using the recursive relation established in Theorem~6 of
\cite{brandes01},
\begin{equation}
  \label{eq:60}
  \delta_s(v) = \sum_{w\in c_s(w)} \frac{\sigma_{s v}}{\sigma_{s w}}
  \; (1 + \delta_s(w)) \text{\;,}
\end{equation}
where $c_s(v)$ denotes the set of child vertices of $v$ in the
directed acyclic graph rooted at $s$.

Finally, in the third stage, we compute the co-betweennesses by
interpreting the relation
\begin{equation}
  \label{eq:70}
  \C(u,v) = \sum_{s\in\V \setminus \{ u,v \}}
  \frac{\delta_s(v)}{\sigma_{s v}} \;
  \sigma_{s v}(u)
\end{equation}
as assigning a contribution of $\frac{\delta_s(v)}{\sigma_{s v}}$
to $C(u,v)$ for each of the $\sigma_{s v}(u)$ shortest 
paths to $v$ that run through $u$.
We accumulate these contributions at each step of the
depth-first traversal when we visit a vertex $v$ by adding
$\frac{\delta_s(v)}{\sigma_{s v}}$ to
$\C(u,v)$ for every ancestor $u$ of the current vertex $v$.

Our proposed algorithms exploit recursions analogous to
those of~\cite{brandes01} to produce run-times that are
in the worst case $O(n_v^3)$, but in empirical
studies were found to vary like $O(n_v n_e+n_v^{2+p}\log
n_v)$ in general, or $O(n_v^{2+p} \log n_v)$ in the case of sparse graphs.
Here $p$ is related to the total number of shortest paths in the
network and seems to lie comfortably between $0.1$ and $0.5$
in our experience.  In the case of unique
shortest paths, it may be shown rigorously that the running time
reduces to $O(n_v n_e+n_v^2 \log n_v)$, and $O(n_v^2 \log n_v)$ if
the network is sparse as well as `small-world' (i.e., with diameter
of size $O(\log n_v)$).  See the appendix for details.

\section{Additional Illustrations}
\label{sec:applications}

We provide in this section additional illustration of the use of 
co-betweenness, based on two other networks graphs.  Both
graphs originally derive from social network analyses in which one goal 
was to understand the flow of certain information among actors.

\subsection{Michael's {\it Strike Network}}
\label{sec:strike.net}

Our first illustration involves the strike dataset of~\cite{michael},
which is also analyzed in detail in Chapter 7 of~\cite{pajek.book}. 
New management took over at a forest products manufacturing facility,
and this management team proposed certain changes to the compensation
package of the workers.  The changes were not accepted by the workers, and
a strike ensued, which was then followed by a halt in
negotiations.  At the request of management, who felt that the
information about their proposed changes was not being communicated
adequately, an outside consultant
analyzed the communication structure among $24$ relevant actors.

\begin{figure} 
  \centering
  \includegraphics[angle=0,scale=0.40]{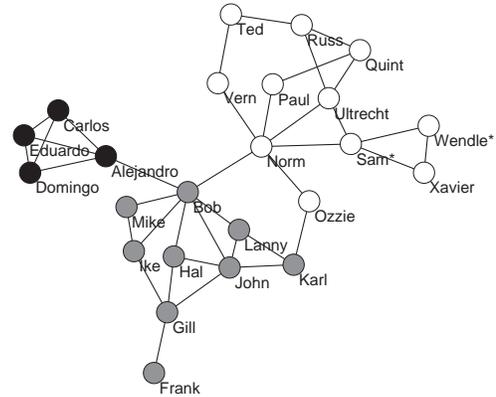}
  \caption{Original strike-group communication network
    of~\cite{michael}. Three subgroups are represented in this
    network: younger, Spanish-speaking employees (black vertices),
    younger, English-speaking employees (gray vertices), and older,
    English-speaking employees (white vertices).  The two union
    negotiators, Sam and Wendle, are indicated by asterix' next to
    their names.  Edges indicate that the two incident actors
    communicated at some minimally sufficient level of frequency
    about the strike.}
\label{fig:strike.group}
\end{figure}

The social network graph in Figure~\ref{fig:strike.group} represents
the communication structure among these actors, with an edge
between two actors indicating that they communicated at some minimally 
sufficient level of frequency about the strike.  
Three subgroups are present in the network: younger, Spanish-speaking
employees (black vertices), younger, English-speaking employees (gray
vertices), and older, English-speaking employees (white vertices).  In
addition, the two union negotiators, Sam and Wendle, are indicated by
asterix' next to their names.  It is these last two that were
responsible for explaining the details of the proposed changes to the
employees.  When the structure of this network was revealed, two
additional actors -- Bob and Norm -- were approached, had the changes
explained to them, which they then discussed with their colleagues,
and within two days the employees requested that their union
representatives re-open negotiations.  The strike was resolved soon
thereafter.

That such a result could follow by targeting Bob and Norm is not
entirely surprising, from the perspective of the network structure.
Both are cut-vertices (i.e., their removal would disconnect the
network), and are incident to edges serving as bridges (i.e., their
removal similarly would disconnect the network) from their respective
groups to at least one of the other groups.  

\begin{figure}  
  \centering
  \includegraphics[angle=0,scale=0.40]{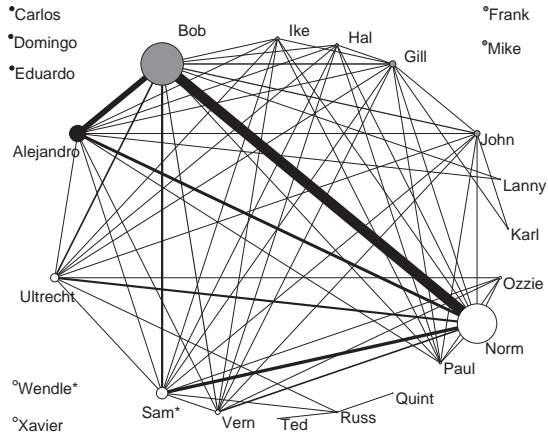}
  \caption{Co-betweenness for the strike-group communication network.
	   Actors located apart from the network, in the corners, 
	   are isolated under this representation, as they have
	   zero betweenness and hence no co-betweenness with any other
	   actors. (Note: Isolated vertices are drawn to have unit
	   diameter, and {\it not} in proportion to their (zero)
	   betweenness.)}
\label{fig:strike.group.CoB}
\end{figure}

Co-betweenness provides a useful alternative characterization, one
which explicitly emphasizes the patterns of communication in the
network, as shown in Figure~\ref{fig:strike.group.CoB}.  
As with Figure~\ref{fig:abilene.CoB}, vertices (now arranged in a
circular layout) are drawn in proportion to their betweenness, and
edges, to their co-betweenness.  Bob and Norm clearly have the largest
betweenness values, followed by Alejandro, who we remark also is a
cut-vertex, but incident to a bridge to a smaller subnetwork than the
other two (i.e., four younger Spanish-speakers, in comparison to nine
younger English-speakers and 11 older English-speakers, for Bob and
Norm, respectively).  The importance of these three actors on the
communication process is evident from the distinct triangle formed by
their large co-betweenness values.  Note that for the two union
representatives, the co-betweenness values suggest that Sam also plays
a non-trivial role in facilitating communication, but that Wendle is
not well-situated in this regard.  In fact, Wendle is not even
connected to the main component of the graph, since his betweenness is
zero (as is also true for six other actors).

\begin{figure}  
  \centering
  \includegraphics[angle=0,scale=0.40]{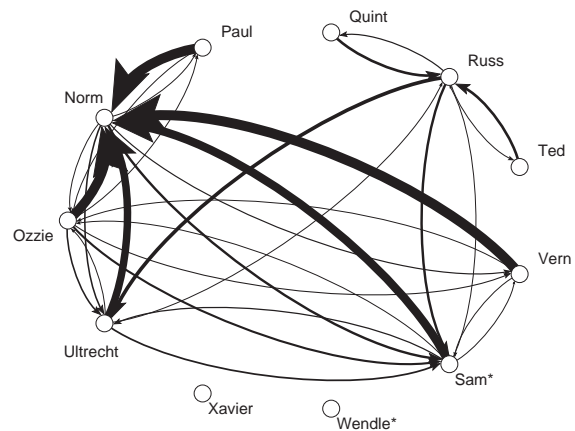}
\caption{Conditional co-betweenness for the older English-speaking
	actors in the strike-group communication network.}
  \label{fig:strike.group.CondP}
\end{figure}

A plot of the standardized co-betweenness $\C^{corr}$ 
shows similar patterns overall, and we have therefore not included it
here.  The conditional betweenness
$\C(u|v)$ for this network 
primarily shows most of the actors with 
large arcs pointing to Bob and Norm, and much smaller arcs pointing
the opposite direction.  This pattern further confirms the influence
that these two actors can have on the other actors in the communication
process.  However, there are also some interesting asymmetrical 
relationships among the actors with smaller parts.  For example,
consider Figure~\ref{fig:strike.group.CondP}, which shows the 
conditional betweenness among the older English-speaking employees.
Ultrecht, for example, clearly has potential for a large amount of control 
on the communication of information passing through Russ, 
and similarly, Karl, on that through John.

\subsection{Zachary's {\it Karate Club} Network}
\label{sec:karate.net}

Our second illustration uses the karate club dataset 
of~\cite{zachary77}.  
Over the course of a couple of years in the 1970s, 
Zachary collected information from the members of a
university karate club, including the number of situations
(both inside and outside of the club) in which interactions
occurred between members. During the course of this study, there
was a dispute between the club's administrator and the principal
karate instructor.
As a result, the club eventually split into two smaller clubs of
approximately equal size---one centered around the administrator
and the other centered around the instructor.

\begin{figure}   
  \centering
  \includegraphics[angle=0,scale=0.40]{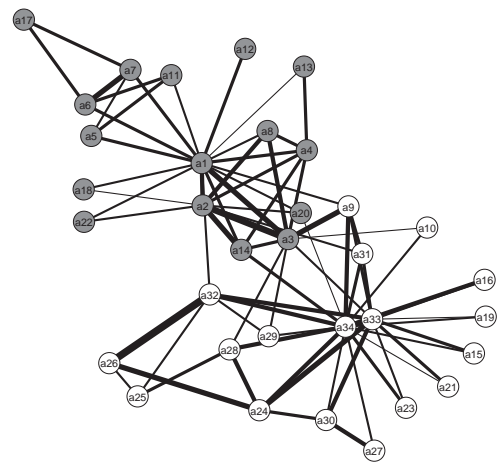}
\caption{Karate club network of~\cite{zachary77}. The gray vertices represent
members of one of the two smaller clubs and the white vertices
represent members who went to the other club. The edges are drawn
with a width proportional to the number of situations in which the
two members interacted.}
\label{fig:zachary.net}
\end{figure}

Figure~\ref{fig:zachary.net} displays the network of social
interactions between club members. 
The gray vertices represent members of one of the two smaller clubs
and the white vertices represent members who went to the other
club. The edges are drawn with a width proportional to the number of
situations in which the two members interacted. The graph clearly
shows that the original club was already polarized into two groups
centered about actors 1 and 34, who were the key players in the
dispute that split the club in two.

\begin{figure}   
  \centering
  \includegraphics[angle=0,scale=0.40]{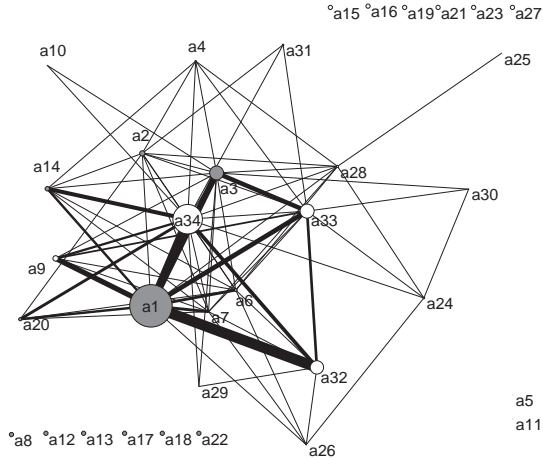}
\caption{Co-betweenness for the karate club network.
	   Actors in the upper-left and lower-right corners, 
	   separated from the connected component, are isolated
	due to zero betweenness.  The two actors in the lower 
	right-hand corner (i.e., $a5$ and $a11$) have non-zero
	betweenness, but are bridges, in the sense that they only serve
	to connect to other vertices, and hence have zero
	co-betweenness.  (Note: The vertices for actors with
	zero betweenness are drawn to have unit diameter, for
	purposes of visibility.) }
\label{fig:zachary.net.CoB}
\end{figure}

The co-betweenness for this network is shown in
Figure~\ref{fig:zachary.net.CoB}.
As in Figure~\ref{fig:zachary.net}, the layout is done using an energy
minimization algorithm.  Again, as in our other examples, the
co-betweenness entries are dominated by a handful of larger values.
As might be expected, actors 1 and 34, who were at the center of the
dispute, have the largest betweenness centralities and are also
involved in the largest co-betweenness'.  More interesting, however,
is the fact that these two actors have a large co-betweenness with
each other -- despite not being directly connected in the original
network graph.  This indicates that they are nevertheless involved in
connecting a large number of other pairs -- probably through key
intermediaries such as actors 3 and 32.  These latter two actors,
while certainly not cut-vertices, nevertheless seem to operate like
conduits between the two groups, quite likely due to their direct ties
to both actor 1 and either of actors 33 and 34, the latter of which
are both central to the group of white vertices.  The co-betweenness
for actors 1 and 32 is in fact the largest in the entire network.

Also of potential interest are the 14 vertices that are isolated from
the network in the co-betweenness representation.  Some of these vertices,
such as actor 8, have strong social interactions with certain other actors
(i.e., with actors 1, 2, 3 and 4), but evidently play a peripheral role in
the communication patterns of the network, as evidenced by their lack of 
betweenness.  Alternatively, there are the vertices like 
those representing actors 5 and 11, who have some betweenness centrality
but nonetheless find themselves cut off from the connected component
in the co-betweenness graph.  An examination of
the definition of the co-betweenness tells us that such vertices
must be bridge-vertices, in the sense that they only serve to 
connect pairs of other vertices, \ie, they only
occur in the middle of paths of length two.

\section{Discussion}
\label{sec:disc}

We introduced in this paper the notion of co-betweenness as a natural
and interpretable metric for quantifying the interplay between pairs
of vertices in a network graph. As we discussed
in different real world examples, this quantity has several interesting features. In 
particular, unlike the usual betweenness centrality which orders the vertices according
to their importance in the information flow on the network, the co-betweenness 
gives additional information about the flow structure and the correlations between 
different actors. Using this quantity, we were able to identify vertices which 
are not the most central ones, but which however play a very important role
in relaying the information and which therefore appear as crucial vertices in 
the control of the information flow.

In principle, of course, one could continue to define higher-order
analogues, involving three or more vertices at a time.  But the
computational requirements associated with calculating such analogues
would soon become burdensome.  In the case of triplets of vertices,
one can expect algorithms analogous to those presented here to scale
no better than $O(n_v^3)$.  Additionally, we remark that, in keeping
with the statistics analogy made in Section~\ref{sec:cob}, it is
likely that the pairwise `correlations' picked up by co-betweenness
captures to a large extent the more important elements of vertex
interplay in the network, with respect to shortest paths.

Following the tendancies in the statistical physics literature on
complex networks~\cite{Barabasi,Vespignani}, it can be of some
interest to explore the statistical properties of co-betweenness in
large-scale networks.  Some work in this direction may be found 
in~\cite{Chua.thesis}, where co-betweenness and functions thereof were
examined in the context of standard network graph models.  The most
striking properties discovered were certain basic scaling relations
with distance between vertices.

On a final note, we point out that, while our discussion here has been
focused on co-betweenness for pairs of vertices in unweighted graphs,
we have also developed the analogous quantities and algorithms for
vertex co-betweenness on weighted graphs and for edge co-betweenness
on unweighted and weighted graphs. Also
see~\cite{chua05:kriging}, where a result is given relating edge
betweenness to the eigen-values of the matrix edge-betweenness
`covariance' matrix, defined in analogy to the matrix $\Omega$ in
Section~\ref{sec:cob}.

\appendix
\label{sec:algorithm}

This appendix contains details specific to the proposed algorithm for
computing co-betweeness, including a derivation of key expressions, a
rough analysis of algorithmic complexity. The pseudo-codes can be
found at the address \cite{address1}. Actual software implementing our
algorithm, written in the {\sc Matlab} software enviroment, is
available at \cite{address2}.

\section{Derivation of Key Expressions}
\label{sec:deriv-key-expr}

Central to our algorithm are the expressions in
\eqref{eq:60} and \eqref{eq:70}, the derivations for which 
we present here.  Before doing so, however, we need to introduce
some definitions and relations.
First note that a simple combinatorial argument will show that
\begin{equation}
  \label{eq:69}
  \sigma_{s t}(v) =
  \begin{cases}
    \sigma_{s v} \, \sigma_{v t} & \text{if $d(s,t) = d(s,v) + d(v,t)$,} \\
    0 & \text{otherwise,}
  \end{cases}
\end{equation}
and 
\begin{equation}
  \label{eq:76}
  \sigma_{s t}(u,v) =
  \begin{cases}
    \sigma_{s u} \, \sigma_{u v} \, \sigma_{v t} & 
    \text{if $d(s,t) = d(s,u)$,} \\
    & + d(u,v) + d(v,t), \\
    \sigma_{s v}\, \sigma_{v u}\, \sigma_{u t} &
    \text{if $d(s,t) = d(s,v)$,} \\
    & + d(v,u) + d(u,t), \\
    0 & \text{otherwise.}
  \end{cases}
\end{equation}
For the the sake of notational simplicity, we will assume, without loss of
generality, that 
\begin{equation}
  \label{eq:77}
  d(s,u) \leq d(s,v).
\end{equation}
for the remainder of this discussion. 

The remaining quantities we need to introduce are notions of the
path-dependency of vertices. In the spirit of \cite{brandes01}, we
define the ``dependency'' of vertices $s$ and $t$ on the vertex
pair $(u,v)$ as
\begin{equation}
  \label{eq:64}
  \delta_{s t}(u,v) = \frac{\sigma_{s t}(u,v)}{\sigma_{s t}} \text{,}
\end{equation}
and we define the dependency of $s$ alone on the pair of vertices
$(u,v)$ as
\begin{equation}
  \label{eq:62}
  \delta_s(u,v) = \sum_{t\in \V \setminus \{ u,v \}} \delta_{s
    t}(u,v)
  = \sum_{t\in \V \setminus \{ u,v \}} \frac{\sigma_{s
      t}(u,v)}{\sigma_{s t}} \text{.}  \text{.}
\end{equation}
Similarly, we define the pair-wise dependency of $s$ and $t$ on a
single vertex $v$ as
\begin{equation}
  \label{eq:80}
  \delta_{s t}(v) = \frac{\sigma_{s t}(v)}{\sigma_{s t}} \text{,}
\end{equation}
and the dependency of $s$ alone on $v$ as
\begin{equation}
  \label{eq:81}
  \delta_s(v) 
  = \sum_{t\in \V \setminus \{ v \}} \delta_{s t}(v) 
  = \sum_{t\in \V \setminus \{ v \}} \frac{\sigma_{s
      t}(v)}{\sigma_{s t}} \text{.} 
\end{equation}
Note that unlike \cite{brandes01}, we exclude $t=v$ from
the sum in \eqref{eq:81}.  Two relations that follow immediately
from these definitions, combined with \eqref{eq:69} and
\eqref{eq:76}, are
\begin{eqnarray}
\nonumber
  \sigma_{s t}(u,v) &=& \sigma_{s u} \, \sigma_{u v} \, \sigma_{v t}\\ 
\nonumber
  &=& \sigma_{s v}(u) \, \sigma_{v t}\\ 
\nonumber
  &=& \frac{\sigma_{s v}(u)}{\sigma_{s v}}\; \sigma_{s v}\,\sigma_{v t}\\
  &=& \delta_{s v}(u) \, \sigma_{s t}(v) \text{,}
  \label{eq:75}
\end{eqnarray}
and 
\begin{equation}
  \label{eq:72}
  \delta_{s t}(u,v) = \frac{\sigma_{s t}(u,v)}{\sigma_{s t}} 
  = \frac{\delta_{s v}(u) \, \sigma_{s t}(v)}{\sigma_{s t}} 
  = \delta_{s v}(u) \, \delta_{s t}(v) \text{.}
\end{equation}
These two relations allow us to show that
\begin{alignat}{2}
  \label{eq:83}
  \delta_s(u,v) &= \sum_{t \in \V \setminus \{ u,v \}} \delta_{s
    t}(u,v) &\qquad& \\ 
  \label{eq:84}
  &= \sum_{t \in \V \setminus \{ u,v \}} \delta_{s v}(u) \,
  \delta_{s t}(v) && \text{by \eqref{eq:72}} \\ 
  \label{eq:85}
  &= \delta_{s v}(u) \, \delta_s(v) 
\end{alignat}
since $\delta_{s u}(v)=0$ by \eqref{eq:77} and using
Eq.~\eqref{eq:75}, we obtain
\begin{equation}
  \label{eq:86}
  \delta_s(u,v)= \frac{\delta_s(v)}{\sigma_{s v}} \; \sigma_{s v}(u) 
\end{equation}
We use this result to re-express the co-betweenness defined in
\eqref{eq:CoB} as 
\begin{align}
  \label{eq:63}
  \C(u,v) &= \sum_{s,t\in \V \setminus \{ u,v \}} \delta_{s t}(u,v) \\
  \label{eq:65}
  &= \sum_{s\in \V\setminus \{ u,v \}} \left(\sum_{t\in
      \V\setminus \{ u,v \}} \delta_{s t}(u,v) \right) \\
  \label{eq:66}
  &= \sum_{s\in\V \setminus \{ u,v \}} \delta_s(u,v) \\
  \label{eq:82}
  &= \sum_{s\in\V \setminus \{ u,v \}}
  \frac{\delta_s(v)}{\sigma_{s v}} \; \sigma_{s v}(u) 
  \text{.}
\end{align}

Lastly, to establish the recursive relation in \eqref{eq:60}, note
that for a child vertex $w\in c_s(v)$ every path to $v$ gives rise
to exactly one path to $w$ by following the edge $(v,w)$.  This
means that
\begin{equation}
  \label{eq:90}
  \sigma_{s w}(v) = \sigma_{s v} \text{\quad for $w\in c_s(v)$,}
\end{equation}
and that 
\begin{equation}
  \label{eq:93}
  \delta_{s w}(v) = \frac{\sigma_{s w}(v)}{\sigma_{s w}} 
  = \frac{\sigma_{s v}}{\sigma_{s w}} \text{\quad for $w \in c_s(v)$.}
\end{equation}
Also note that for $t=w$ we have
\begin{equation}
  \label{eq:91}
  \delta_{s t}(w) = 1\text{.}
\end{equation}
This allows us to decompose $\delta_s(v)$ in essentially the
same manner as \cite{brandes01}, namely,
\begin{alignat}{2}
  \label{eq:68}
  \delta_s(v) &= \sum_{t\in \V \setminus \{v\}} \delta_{s t}(v) \\
  \label{eq:78}
  &= \sum_{t\in \V \setminus \{v\}} \; \sum_{w\in c_s(v)} \delta_{s t}(v,w) \\
  \label{eq:79}
  &= \sum_{w\in c_s(v)} \; \sum_{t\in \V \setminus \{v\}} \delta_{s t}(v,w) \\
  \label{eq:88}
  &= \sum_{w\in c_s(v)} \; \sum_{t\in \V \setminus \{v\}} \delta_{s w}(v) \,
  \delta_{s t}(w) &\qquad&\text{by \eqref{eq:72}}\\
  \label{eq:87}
  &= \sum_{w\in c_s(v)} \frac{\sigma_{s v}}{\sigma_{s w}}  \left( 
    1 + \sum_{t\in \V\setminus \{ v,w \}} 
    \delta_{s t}(w) \right)
\end{alignat}
Using \eqref{eq:93} and \eqref{eq:91}, we then obtain
\begin{equation}
  \label{eq:89}
  \delta_s(v)= \sum_{w\in c_s(v)} \frac{\sigma_{s v}}{\sigma_{s w}} \; (1 +
  \delta_s(w)) \text{.}
\end{equation}
Where the last equality is due to the fact that since $w$ is a
child of $v$ we have $\sigma_{s v}(w) = 0$ and thus $\delta_{s
  v}(w)=0$. 

\section{Algorithmic Complexity}
\label{sec:algor-compl}

Standard breadth-first search results put the running time for the
first stage of our algorithm at $O(n_e)$, and since we touch each edge
at most twice when we compute the dependency scores $\delta_s(v)$, the
running time for the second stage is also $O(n_e)$.  Since we repeat
each stage for each vertex in the network, the first two stages have a
running time of $O(n_v n_e)$.  The running time for the depth-first
traversal, that occurs during the third stage, depends on the number
and length of all shortest paths in the network.  Overall, we visit
every shortest path once and compute a co-betweenness contribution for
each edge of every shortest path. For `small-world' networks i.e.,
networks with an $O(\log n_v)$ diameter, we must compute
$O(\sigma\cdot\log n_v)$ contributions, where $\sigma =
\sum_{u,v\in\V} \sigma_{u v}$ is the total number of shortest paths in
the network. So the overall running time for the algorithm is $O(n_v
n_e+\sigma\log n_v)$. Empirical evidence
suggests that the upper bound for the average $\frac{1}{\abs{\V}}\sum_{u\in\V}
\sigma_{u v}$ ranges from $n_v^{0.19}$ to $n_v^{0.32}$ for 
common random graph models, and at worst has been seen to reach
$n_v^{0.62}$ in the case of a network of airports. (In the latter
case, there were extreme fluctuations in
$\frac{1}{\abs{\V}}\sum_{u\in\V} \sigma_{u v}$ so the total number
of shortest paths, $\sigma$, might be much smaller than
$n_v(n_v-1)$ times this upper bound.) This suggests a running time
of $O(n_v n_e + n_v^{2+p} \log n_v)$, though it is an open
question to show this rigorously.  In the case of sparse networks,
where $n_e \sim n_v$, this reduces to a running time of
$O(n_v^{2+p} \log n_v)$.

\end{document}